\documentclass[12pt,a4paper]{article}

\usepackage{amsmath}                    % From AMS/CTAN distribution
\usepackage{amssymb}                    % From AMS/CTAN distribution
\usepackage{amsthm}                     % From AMS/CTAN distribution
\usepackage{amsfonts}                   % From AMS/CTAN distribution
\usepackage[pdftex]{graphicx} 
\usepackage{color}
\usepackage{bbm}

\theoremstyle{definition}
\newtheorem{dfn}{Definition}
\theoremstyle{plain}
\newtheorem{thm}[dfn]{Theorem}

\newtheorem{lemma}[dfn]{Lemma}

\newenvironment{prf}{\noindent {\bf Proof.}}{\begin{flushright}\vspace{-2em}\footnotesize$\boxdot$\normalsize\end{flushright}\smallskip}
\begin{document}

\title{Graphic Realizations of Joint-Degree Matrices\footnote{All the results of this work are as of July 6, 2009. Throughout the paper, however, references to more recent works have been added, so that it is up to date.}}

\author{Georgios Amanatidis\thanks{Department of Informatics, Athens University of Economics and Business, Email: \textsf{gamana@aueb.gr}.
Supported in part by an ACO and an ARC 
Georgia Tech Fellowiship, and NSF-CCF-TF-0830683.}
\and Bradley Green\thanks{Facebook, Email: \textsf{Brg@fb.com}.
Supported in part by an NSF VIGRE Fellowship.} 
\and Milena Mihail\thanks{College of Computing, 
Georgia Institute of Technology, Email: \textsf{mihail@cc.gatech.edu}. 
Supported in part by NSF-CCF-0539972 and NSF-CCF-TF-0830683.}}

%\date{July 6, 2009$\S$}
\date{\vspace{-5ex}}

\maketitle
\thispagestyle{empty} 

\hspace{1cm}

\begin{abstract}
In this paper we introduce extensions and modifications of the
classical degree sequence graphic realization problem studied by Erd\H{o}s-Gallai and Havel-Hakimi, 
as well as of the corresponding connected graphic realization version.
We define the joint-degree matrix graphic 
(resp. connected graphic) realization problem,
where in addition to the degree sequence, 
the exact number of desired edges 
between vertices of different degree classes
is also specified. 
We give necessary and sufficient conditions, and polynomial time decision and construction algorithms 
for the graphic and connected graphic realization problems.
These problems arise naturally in the current topic of graph modeling for complex networks.
From the technical point of view, 
the joint-degree matrix realization algorithm is straightforward.
However, the connected joint-degree matrix realization algorithm 
involves a novel recursive search of suitable local graph modifications.
Also, we outline directions for further work of both theoretical and practical interest. In particular,
we give a Markov chain which converges to the uniform distribution over all realizations. We show
that the underline state space is connected, and we leave the question of the mixing rate open.
%
%
%\vspace{20 ex}  \noindent$^\S\,$All the results of this work are as of July 6, 2009. Throughout the paper, however, references to more recent works have been added, so that it is up to date.
\end{abstract}

%\newpage
%\setcounter{page}{1}
\section{Introduction}
\label{sec:introduction}

Let $d_1 \geq d_2 \geq \ldots \geq d_n$ be a sequence of integers.
The classical {\it graphic realization} problem asks
if there a simple graph on $n$ vertices
whose degrees are exactly $d_1 \geq d_2 \geq \ldots \geq d_n$.
Erd\H{o}s and Gallai
showed that the natural necessary conditions for graphic realizability,
namely that each subset of the highest $k$ degree vertices can absorb
their degrees within their subset and the degrees of the remaining vertices:
$\sum_{i=1}^k d_i \leq k(k\! - \! 1) + \sum_{i=k+1}^n \min \{ k,d_i \} $,
are also sufficient \cite{ErdosGallai,bergebook}.
The well known Havel-Hakimi
algorithm \cite{HavelHakimi1,HavelHakimi2} achieves
a realization
in an efficient greedy way. It repeatedly sorts the vertices according
to residual unsatisfied degree, picks any vertex
of residual degree $d_i$,
and connects it to the $d_i$ vertices of highest residual degree.
The process is repeated until all the degrees are satisfied.
If one further wants to construct a {\it connected graphic realization}
(a requirement which is clearly important in networking),
Erd\H{o}s and Gallai showed that
the obvious necessary condition $\sum_{i=1}^n d_i \geq 2(n\! - \! 1)$
(i.e., there is a spanning tree) is also sufficient.
In particular, it is easy to see that a non-connected realization
can be transformed to a connected
realization by a sequence of flips, each flip breaking a cycle inside
a connected component, and reducing the number of
connected components by one.
A  ``flip''
picks two edges  $xy$ and $uv$
such that $xu$ and $yv$ are not edges,
removes $xy$ and $uv$ from the graph,
and adds $xu$ and $yv$ to the graph.
It is clear that flips do
not change the degrees of the graph.

Now, let $V \! = \! [n]$ be a set of vertices.
Let ${\mathbb V} \! = \! \{ V_1 , V_2 , \ldots , V_k \}$
be a partition of $V$ denoting
subsets of vertices with the same degree and let $d:\mathbb V \rightarrow {\mathbb N}$
be a function denoting the degree
of vertices in class $V_i$.
Let $D \! = \! (d_{ij})$ be a $k\times k$ matrix
denoting the number of edges between $V_i$ and $V_j$;
if $i\! = \! j$ it is the number of edges entirely within $V_i$.
The {\it joint-degree matrix graphic realization} problem is,
given $\langle{\mathbb V},d,D\rangle$, decide
whether there is a simple graph $G$ on $V$, such that,
$\forall i$ each vertex in $V_i$ has degree $d(V_i)$, 
$\forall i \neq j$ there are exactly $d_{ij}$ edges between $V_i$ and $V_j$,
and, $\forall i$, there are exactly $d_{ii}$ edges entirely inside $V_i$.
The {\it joint-degree matrix connected graphic realization} problem
is to decide whether a \emph{connected} graphic realization for 
$\langle{\mathbb V},d,D\rangle$
exists. Furthermore, we want to either construct such a realization,
or output a certificate that no such graphic connected realization 
exists.
In this paper we give necessary and sufficient conditions, and polynomial (in $n$) time algorithms
for the decision and construction of the joint-degree graphic
and connected graphic realization problems.

The practical significance of joint-degree
matrix realization problems arises in graph
models for several classes of complex networks.
For instance, in networking, models for Internet
topologies are constantly used to simulate
network protocols and predict network evolution.
Commonly used topology generators, such as
GT-ITM \cite{Zegura2,Zegura3} which generates random nearly regular graphs,
and power-law based generators \cite{Faloutsos,ACL1,ACL2,Brite2,Inet,ALENEX03}, generate random graphs.
However, the properties of topologies constructed using
purely random graph models were challenged,
most notably in \cite{hot2,alderson}.
Using qualitative arguments and striking images,
they argued that 
power law random graphs construct a dense core
of nodes with very high degrees,
while nodes of smaller degree are mostly attached to the periphery of the
network. On the other hand, highly optimized Internet topologies
place low degree but very high bandwidth routers at the center of the network,
while high degree nodes are mostly placed in the periphery
to split the signal manyways toward the end users. 
To quantify their argument, \cite{alderson} used a random graph 
$G_{\rm R}(V,E_{\rm R})$
under several power law 
models \cite{nsbook2,nsbook3,chungbook2,newmanbook,durrettbook},
and a real network topology $G(V,E)$.
They found that
$\sum_{uv \in E_{\rm R}} {\rm deg}(u) {\rm deg}(v)$  
is much larger than
$\sum_{uv \in E} {\rm deg}(u) {\rm deg}(v)$.
Independently, \cite{Newman1,Newman5} made the same observation
for several other technological and biological networks.

Going one step further, \cite{UCSDnet1,UCSDnet2}
argued that, 
a determining metric
for a graph of given degrees
to resemble a real network topology,
is the specific number of links
between vertices in different degree classes.
Using heuristics that presumably approximate
the target number of edges between degree classes, 
\cite{UCSDnet2} constructed graphs
strikingly similar to real network topologies. 
The joint-degree matrix graphic and connected graphic realization problems 
studied in Sections \ref{sec:algorithm1} and \ref{sec:algorithm2}
formalize the approach of \cite{UCSDnet1,UCSDnet2}. In general, one would want 
to construct a uniformly random realization of $\langle{\mathbb V},d,D\rangle$.
We state this as an open problem in Section 4. However, for practical purposes, 
the heuristics of \cite{UCSDnet1,UCSDnet2} achieved very satisfactory results 
using randomness in a configuration model adjusted to the $\langle{\mathbb V},d,D\rangle$ problem.
On the other hand, there are no theoretical results concerning the properties of this model.
In the same sense, our joint degree matrix realization algorithm in Section 2, also
allows substantial randomness in the choice of the edge to be added in each of its greedy steps.

In Section \ref{sec:algorithm1} we address the joint-degree matrix
graphic realizability problem. We show that the natural necessary conditions
are also sufficient, and can be checked efficiently. 
We also obtain a second polynomial time construction 
algorithm, which is a key for the algorithm in Section \ref{sec:algorithm2}.
This Balanced Degree Algorithm constructs the graph by increasing the number of
edges, without increasing the number of connected components.
In Section \ref{sec:algorithm2} we address 
the joint-degree matrix connected graphic realizability problem.
By sharp contrast to the degree sequence connected realization, 
here the necessary and sufficient conditions are fairly complex,
and of exponential size. However, using a recursive algorithm
that searches for suitable local graph modifications to construct
a connected graph, we manage to either construct such a graph in polynomial time, or identify 
at least one necessary condition that fails to hold. 
In Section \ref{sec:discussion} we discuss 
structural differences between degree sequence 
and joint-degree matrix problems. 
In particular, the former are known to be related to 
matchings, while no corresponding fact is known for the latter. 
Finally, in Section \ref{sec:discussion}, we propose a natural Markov chain 
for sampling from $\langle{\mathbb V},d,D\rangle$, and we show  that it is 
ergodic.

\noindent \textbf{Recent related work.} Independently, \cite{SP12, CDEM15, GTM15} give polynomial time algorithms for constructing a graph in $\langle{\mathbb V},d,D\rangle$. Moreover, in \cite{SP12} an alternative proof is proposed for the fact that the Markov chain we define in Section \ref{sec:discussion} is ergodic. This proof however was flawed, as noted in \cite{CDEM15}, where an alternative proof is given as well. With respect to the mixing time of this Markov chain, \cite{SP12} performed experiments based on the autocorrelation of each edge; these experiments
suggest that the Markov chain mixes quickly. In a more recent work, \cite{EMT15} shows fast mixing for a related Markov chain over the subset of $\langle{\mathbb V},d,D\rangle$ that contains the balanced realizations, i.e. realizations where for each $i, j$ the edges connecting $V_i$ to $V_j$ are as uniformly distributed on $V_i$ as possible. (Notice that this is not what we call a \emph{balanced} graph, e.g. in Lemma \ref{lem:last} or in Section \ref{sec:algorithm2}.)

%==========================================================
%==========================================================

\section{Joint-Degree Matrix Graphic Realization}
\label{sec:algorithm1}

%==========================================================
%==========================================================

Let $V \! = \! [n]$ be a set of vertices and 
${\mathbb V} \! = \! \{ V_1 , V_2 , \ldots , V_k \}$
be a partition of $V$ denoting
subsets of vertices with the same degree. Let $d:\mathbb V \rightarrow {\mathbb N}$
be a function denoting the degrees
of vertices in class $V_i$ and $D \! = \! (d_{ij})$ be a $k\times k$ matrix
denoting the number of edges between $V_i$ and $V_j$;
if $i\! = \! j$ it is the number of edges entirely within $V_i$.
The {\it joint-degree matrix graphic realization} problem is,
given $\langle{\mathbb V},d,D\rangle$, decide
whether there is a simple graph $G$ on $V$, such that,
$\forall i$ each vertex in $V_i$ has degree $d(V_i)$,
$\forall i \neq j$ there are exactly $d_{ij}$ edges between $V_i$ and $V_j$,
and $\forall i$, there are exactly $d_{ii}$ edges entirely inside $V_i$.
We use the notation $\langle{\mathbb V},d,D\rangle$
to also denote the set of all such graphs.

We will prove that the following natural
necessary conditions for the instance $\langle{\mathbb V},d,D\rangle$
to have a graphic realization are also sufficient:\smallskip\\
{\it (i) Degree feasibility}:
$2d_{ii} \! + \! \sum_{j \in [k],j\neq i} d_{ij} =
|V_i| \cdot d(V_i)$, for $ 1 \leq i \leq k$. \smallskip\\
{\it (ii) Matrix feasibility}:
The matrix $D$ is symmetric with nonnegative integral entries,
and
$d_{ij} \leq |V_i|\cdot|V_j|$, for $ 1 \leq i < j \leq k$,
while
$d_{ii} \leq |V_i|\cdot(|V_i|\! - \! 1)/2$, for $ 1 \leq i \leq k$.\smallskip

There is a straightforward algorithm for constructing a graph 
$G \in \langle{\mathbb V},d,D\rangle$. First, the algorithm
constructs a graph $G'$ that has the ``right'' number of edges
between any $V_i$, $V_j$ (or within a $V_i$). Then, the degrees 
within each $V_i$ are taken care of, resulting in a graph
$G \in \langle{\mathbb V},d,D\rangle$.

The algorithm proceeds as follows:\medskip\\
Start with an empty graph $G$ on $V$\\
For each $i$\\ $~~~~$ choose arbitrarily $d_{ii}$ edges between
vertices of $V_i$, and add them to $G$\\
For each $i,j$\\ $~~~~$ choose arbitrarily $d_{ij}$ edges between
vertices of $V_i$ and $V_j$, and add them to $G$\\
For each $i$\\ 
$~~~~$ While not all degrees in $V_i$ are equal\\
$~~~~$ $~~~$ Choose $u,v \in V_i$ such that ${\rm deg}(u)< d(V_i)$
and ${\rm deg}(v)> d(V_i)$\\
$~~~~$ $~~~$ Find $\min \{|{\rm deg}(u)-d(V_i)|,|{\rm deg}(v)-d(V_i)|\}$ 
neighbors of $v$ that are not neighbors of $u$\\
$~~~~$ $~~~$ Disconnect them from $v$ and connect them to $u$\\
Output $G$\medskip

To see that the algorithm works, first notice that if 
$d_{ij} \leq |V_i|\cdot|V_j|$, for $ 1 \leq i < j \leq k$,
and
$d_{ii} \leq |V_i|\cdot(|V_i|\! - \! 1)/2$, for $ 1 \leq i \leq k$
the edge-adding part of the algorithm works. This results in a graph 
$G$ that satisfies the $d_{ij}$ requirements, but not necessarily 
the degree requirements. 

Now, assume that there exist some $i$ such that
not all the degrees in $V_i$ are equal to $d(V_i)$. If 
$2d_{ii} \! + \! \sum_{j \in [k],j\neq i} d_{ij} =
|V_i| \cdot d(V_i)$, this means that there exist 
$u,v \in V_i$ such that ${\rm deg}(u)< d(V_i)$
and ${\rm deg}(v)> d(V_i)$. Also, there are $|{\rm deg}(u)-{\rm deg}(v)|$
neighbors of $v$ that are non-neighbors of $u$, and 
$|{\rm deg}(u)-{\rm deg}(v)|>\min \{|{\rm deg}(u)-d(V_i)|,|{\rm deg}(v)-d(V_i)|\}$.
Also, notice that each iteration in the ``while" loop reduces 
the number of ``wrong degrees" by at least one, without affecting 
the $d_{ij}$ requirements. That is, in at most $n$ iterations
$G \in \langle{\mathbb V},d,D\rangle$.\medskip

Although the Joint-Degree Matrix Graphic Realization problem has a
straightforward solution, this is not the case if we also ask for the resulting 
graph to be connected. Before we move to this problem, we  present an 
alternative algorithm for the Joint-Degree Matrix Graphic Realization
that we will need later.

%\vspace{-10pt}
%==========================================================
%==========================================================

\subsection*{Balanced Degree Algorithm}

%==========================================================
%==========================================================

\label{sec:BalancedDegreeAlgorithm}

This construction algorithm grows the graph $G$
in iterations, one edge at a time, starting from the empty graph $G_0$,
keeping the edges between each $V_i$ and $V_j$ always $\leq$ $d_{ij}$
(resp. the edges within each $V_i$ always $\leq$ $d_{ii}$),
and ending with a realization $G \in
\langle{\mathbb V},d,D\rangle$.
The key idea of the algorithm is to
maintain a {\it balanced degree invariant}
within each $V_i$.
If $G_{\ell}$ is the graph after iteration ${\ell}$,
the algorithm maintains 
$\max_{v\in V_i} \deg_{G_{\ell}}(v) - 
\min_{v\in V_i} \deg_{G_{\ell}}(v) \leq 1$,
for $1 \leq i \leq k$ (where $\deg_G(v)$ is the
degree of vertex $v$ in graph $G$, as usual).
This motivates the following definition.
For the graph $G_{\ell}$ after iteration $\ell$,
for all $1 \leq i \leq k$, 
let $N_{i,G_{\ell}} \! =  \! \{ v\in V_i :
\deg_{G_{\ell}}(v) = \min_{u\in V_i} \deg_{G_{\ell}}(u)  \}$
and let $M_{i,G_{\ell}} \! =  \! \{ v\in V_i :
\deg_{G_{\ell}}(v) = \max_{u\in V_i} \deg_{G_{\ell}}(u)  \}$.

The algorithm proceeds as follows: While there is some $i$ and $j$ (possibly $i=j$)
such that $d_{ij}$ is not satisfied,
the construction algorithm picks any such $i$ and $j$
and adds an edge between $V_i$ and $V_j$ (resp. inside $V_i$),
while maintaining the balanced degree invariant
and without affecting the extend to which the other $d_{uv}$'s
are satisfied.
Let us assume that we are at the beginning of the $({\ell}\! + \! 1)$th
iteration, and $i,j$ have been picked such that $d_{ij}$ is not satisfied.
Let $G = G_\ell$ and assume $N_i=N_{i,G_{\ell}}$ and $M_i=M_{i,G_{\ell}}$ are suitably defined for all $i$. 
There are several cases to consider:\smallskip \\
If $i \neq j$ consider Cases A1, A2 and A3 below, 
in the order that they are listed:\\
{\bf Case A1:} if there exist $u\in N_i, v\in N_j$
such that $uv \notin E(G)$ then add $uv$ to $E(G)$;\\
{\bf Case A2} if there exist $u\in N_i, v\in M_j$ 
such that $uv \notin E(G)$ then\\
$~~~~~~$pick a $v^{\prime} \in N_j$ and find a neighbor $x$ of $v$ 
such that $v^{\prime}x \notin E(G)$;\\
$~~~~~~$delete the edge $vx$ from $E(G)$
and add the edges $uv$ and $v^{\prime}x$ to $E(G)$;\\
{\bf Case A$\mathbf 2^{\:\prime}$:} if there exist $u\in M_i, v\in N_j$
such that $uv \notin E(G)$ then symmetric to Case A2;\\
{\bf Case A3:} find $u\in M_i, v\in M_j$ 
such that $uv \notin E(G)$;\\
$~~~~~~$pick $u^{\prime} \in N_i$ and find a neighbor $x$ of $u$ 
such that $u^{\prime}x \notin E(G)$;\\
$~~~~~~$pick $v^{\prime} \in N_j$ and find a neighbor $y$ of $v$
such that $v^{\prime}y \notin E(G)$;\\
$~~~~~~$delete the edges $ux,vy$ from $E(G)$ 
and add the edges $u^{\prime}x,uv,u^{\prime}y$ to $E(G)$;\smallskip\\
If $i = j$ consider Cases B1, B2 and B3 below,
in the order that they are listed:\\
{\bf Case B1:} if there exist $u , v\in N_i$
such that $uv \notin E(G)$ then add $uv$ to $E(G)$;\\
{\bf Case B2} if there exist $u\in N_i, v\in M_j$
such that $uv \notin E(G)$ then\\
$~~~~~~$if $|N_i|= 1$ then add $uv$ to $E(G)$\\
$~~~~~~$elseif $|N_i|>1$ then\\
$~~~~~~~~~~~~~$pick a $v^{\prime} \in N_i$ and find a neighbor $x$ of $v$
such that $v^{\prime}x \notin E(G)$;\\
$~~~~~~~~~~~~~$delete the edge $vx$ from $E(G)$
and add the edges $v^{\prime}x$ and $uv$ to $E(G)$;\\
{\bf Case B3:} find $u,v\in M_i$
such that $uv \notin E(G)$;\\
$~~~~~~$if $|N_i| = 1$ then\\
$~~~~~~~~~~~~~$pick a $w \in N_i$ and find a neighbor $x$ of $u$
such that $wx \notin E(G)$;\\
$~~~~~~~~~~~~~$delete the edge $ux$ from $E(G)$
and add the edges $wx$ and $uv$ to $E(G)$;\\
$~~~~~~$elseif $|N_i|>1$ then\\
$~~~~~~~~~~~~~$pick $w,w^{\prime} \in N_i$;\\
$~~~~~~~~~~~~~$find a neighbor $x$ of $u$ such that $wx \notin E(G)$;\\
$~~~~~~~~~~~~~$find a neighbor $y$ of $v$ such that $w^{\prime}y\notin E(G)$;\\
$~~~~~~~~~~~~~$delete the edges $ux,vy$ from $E(G)$
and add the edges $wx,uv,w^{\prime}y$ to $E(G)$;

\begin{thm}
\label{them:algorithm1}
If the degree and matrix feasibility conditions hold,
then the above algorithm  
constructs a graph $G\in \langle{\mathbb V},d,D\rangle$.
The algorithm runs in time polynomial in $n$. 
In particular, $\forall {\ell}$, if $G_{\ell}$ is the graph at the end of the  ${\ell}$th 
iteration, 
then, the number of edges between $V_i$ and $V_j$ (resp. inside $V_i$) 
have increased by one, the number of edges between and inside all other 
degree classes have not changed,
and the balanced degree invariant holds.
\end{thm}

%\noindent The proof of Theorem \ref{them:algorithm1} is in the Appendix.\\
\begin{prf}
Assume that we are at the beginning of the ${\ell}$th iteration and 
the balanced degree invariant holds.  
We show that the above algorithm maintains 
the invariant after the next edge is added.

Observe first that, for all $i$,
the sets $N_i$ and $M_i$ are always nonempty and 
$M_i\cup N_i=V_i$. In fact, either $N_i=M_i=V_i$, 
or $\{M_i,N_i\}$ is a partition of $V_i$. 
In the sequel, if we refer to $M_i$, then we are assuming that $N_i \ne M_i$.

Now let $i$ and $j$ be two indices so that $G_{\ell}$ has less 
than $d_{ij}$ edges in the subgraph induced by $V_i \cup V_j$.  
By matrix feasibility, 
there is an edge $uv \not \in E(G_{\ell})$ with $u \in V_i$ and $v \in V_j$.

Consider first the case $i \neq j$. 
If the edge $uv$ falls into Case A1, 
the invariant clearly holds for $G_{\ell} + uv$.  
In Case A2 (A$2'$ is symmetric) since $M_j \ne N_j$, 
there is a $v' \ne v \in N_j$.  
Since $\deg(v') < \deg(v)$, there is a neighbor $x$ of $v$ 
such that $v'x \not \in G_{\ell}$.  
The specified actions maintain the invariant.

If Case A3 is reached, we must have $uv$ such that $u \in M_i$ and $v \in M_j$. 
Consider $u^{\prime}\in N_i, v^{\prime}\in N_j$.
Since no edge in Case A2 (or A$2'$) was available, 
we must have $uv^{\prime}, u^{\prime}v$ and $u^{\prime}v^{\prime} \in E(G_{\ell})$.
Since $v$ is a neighbor of $u^{\prime}$ but not of $u$,
and $\deg_G(u^{\prime})\leq \deg_G(u)$, 
there must exist some $x$ that is a neighbor of $u$ but not of $u^{\prime}$. 
Similarly, there must exist some $y$ that is a neighbor of $v$ 
but not of $v^{\prime}$ (possibly $y=x$). 
Because $u,u^{\prime}$ are both in $V_i$ 
and $v,v^{\prime}$ are both in $V_j$,
we can remove $xu, yv$ and add $xu^{\prime}, yv^{\prime}$. 
This way, the extend to which the requirements of matrix $D$
are satisfied is not affected, 
but the degrees of $u, u^{\prime}, v$ and $v^{\prime}$ change
so that we can add $uv$ to $G$ and the statement of the theorem is true.

Next consider the case $i=j$. 
If $u,v \in N_i$, then it is clear that Case B1 maintains the invariant.

Now, assume there is no available edge with both ends in $N_i$, 
but there exist $u\in M_i, v\in N_i$ such that $uv\notin E(G)$.
If $|N_i|=1$, adding $uv$ to the current graph maintains 
the invariant since after the addition 
$M_i = \left\{ v \right\}$, $N_i = V_i - \left\{ v \right\}$.

So suppose that $|N_i|\geq 2$, and let $v' \ne u \in N_i$. 
Since $\deg(v) > \deg(v')$, there exists an edge $vx \in G_{\ell}$ such that 
$v'x \not \in G_{\ell}$.  Note that $x \ne u$. 
Now the specified actions satisfy the theorem.

The last possibility is that the only available 
edges have $u,v \in M_i \ne N_i$.
Notice that by exhausting Case B2 first, 
$w\in N_i$ implies $wu, wv \in E(G_{\ell})$.
Again, we consider two cases: $|N_i|=1$ and $|N_i|\geq 2$.

In the former case, pick a $w\in N_i$. Since $v$ is a 
neighbor of $w$ but not of $u$, 
and $deg_G(w) < deg_G(u)$, there must exist some $x$ that is a 
neighbor of $u$ but not of $w$. 
Because $u,w$ both are in $V_i$  
we can remove $xu$ and add $xw$. 
This way the $d_{ij}$ requirements are not affected for any $j$, 
but the degrees of $u$ and $w$ change so that we can add $uv$ to $G$ and keep
the invariant true.

In the latter case, where $|N_i|\geq 2$, 
pick a $w, w'\in N_i$. Notice that $w$
is a neighbor of $w'$ 
(or else $ww'$ would have been added in B1). 
Since $v$ is a neighbor of $w$ but not of $u$,
and $deg_G(w)\leq deg_G(u)$, 
there must exist some $x$ that is a neighbor of $u$ but not of $w$. 
Similarly, there must exist some $y$ that is a neighbor of $v$ 
but not of $w'$ (possibly $y=x$). 
Because $u,v,w,w'$ all are in $V_i$, we can remove $xu, yv$ 
and add $xw, yw'$. This way the $d_{ij}$ requirements 
are not affected for any $j$, but the degrees of $u,v, w$ and $w'$ change 
so that we can add $uv$ to $G$ and the invariant holds.

As long as there exists some $d_{ij}$ not yet satisfied, 
the algorithm manages to increase the number of $V_i$--$V_j$ edges
by one without changing the number of $V_{i'}$--$V_{j'}$ edges, 
for any $\{i',j'\}\neq\{i,j\}$. Therefore in $m=\sum_{i\leq j} d_{i,j}$
iterations, all the edge requirements are met. 
Now, since degree feasibility holds and for any $i$ 
we have $2d_{ii}+\sum_{j\neq i} d_{ij} = |V_i| \cdot {d}(V_i)$, 
we get that $\forall v\in V_i, \ deg_{G}(v)={d}(V_i)$, as desired.
\end{prf}

\noindent{\bf Remark 1:} The transformations of  
adding and deleting edges in all non-trivial cases of the algorithm
resemble augmenting paths.
However, in general, these transformation are not
augmenting paths. For example, in Case A3,
the sequence of edges
$u^{\prime}x$, $xu$, $uv$,
$vy$ and $yv^{\prime}$
includes the case where $x \! = \! y$.
We clearly have an alternating sequence but not a path.
We shall revisit this comment in Section \ref{sec:discussion}.\medskip

\noindent{\bf Remark 2:} We claim that the construction algorithm
never increases the number of connected components.
In particular, it can be verified that, in all cases, when
an edge is removed, a path between its endpoints is created 
by the edges added in the same iteration.
In particular, if the graph $G_{\ell}$ at iteration $\ell$ is connected,
the final output graph $G$ will be connected.
We will use this fact critically in the algorithm
which constructs a connected realization of $\langle{\mathbb V},d,D\rangle$ 
in Section \ref{sec:algorithm2}.

%==========================================================
%==========================================================

%\vspace{-10pt}
\subsection*{A generalization}
\label{sec:Generalizations}

%==========================================================
%==========================================================

It is natural to consider the generalization of the joint degree matrix problem 
$\langle \mathbb V, d, D^*\rangle$, where we allow 
the entries of $D^*$ to be in ${\mathbb N_0} \cup \{ * \}$. 
If $d_{ij} \! = \! *$, there is no restriction on the number of edges
between the corresponding sets.
We can use the 
main idea of the proof of Theorem \ref{them:algorithm1}
to provide a polynomial time construction algorithm.
This is proved in Theorem \ref{thm:*gener} below.

Notice that, if the entire matrix $D^*$ consists of $*$'s,
then this is the standard degree sequence realizability problem.
In the case where $D^*$ contains both integers and $*$'s, if $\langle \mathbb V, d, D^*\rangle$ is nonempty, 
then there exists some graph
$G\in\langle \mathbb V, d, D^*\rangle$ such that the subgraph $H$ of $G$ defined
by the integer entries of $D^*$ satisfies the balanced degree invariant. This is proved in Lemma \ref{lem:last} below.
We call such a graph $G$
a \emph{balanced $\langle \mathbb V, d, D^*\rangle$ graph}.

\begin{lemma}
\label{lem:last}
If $\langle \mathbb V, d, D^*\rangle\neq \emptyset$, then there exists a
balanced $\langle \mathbb V, d, D^*\rangle$ graph.
\end {lemma}
\begin{prf}
Let $G\in \langle \mathbb V, d, D^*\rangle$ and consider the subgraph $H$ of $G$
with vertex set $V(H)=V(G)$ and edge set
$E(H)=\{uv\in E(G)\::\: u\in V_i, v\in V_j,\: d_{ij}\neq *\}$. Assume that $H$
does not satisfy the balanced degree invariant. 
Then consider $G'$ on $V(G)$ with edge set
$E(G')=E(G)\setminus E(H)$. We can find $u$ and $v$ in some $V_i$ such that,
$\deg_{H}(u)> \deg_{H}(v) + 1$, and thus $\deg_{G'}(u)< \deg_{G'}(v) -1 $. 
We can pick
a neighbor $x$ of $u$ in $H$ that is not a neighbor of $v$, 
and a neighbor $y$
of $v$ in $G'$ that is not a neighbor of $u$. 
We remove $ux, vy$ in $G$ and add
$vx$ and $uy$. 
We repeat the above procedure until $H$ satisfies the balanced
degree invariant. 
Notice that the edge flips are such that the resulting graph
$G$ is still in $\langle \mathbb V, d, D^*\rangle$.
\end{prf}

\begin{thm}\label{thm:*gener}
If $\langle \mathbb V, d, D^*\rangle\neq \emptyset$, then we can construct a graph $\langle \mathbb V, d, D^*\rangle$
in polynomial time.
\end{thm}
\begin{prf} Notice that for any balanced graph $G\in \langle \mathbb V, d, D^*\rangle$,
if we consider the subgraph $\bar{H}$ defined by the $*$ entries of
$D^*$, it is a realization
of the same degree sequence, 
say $d_1',d_2',\ldots,d_n'$ and the same
edge restrictions, i.e.,
there are no edges between vertices of $V_i$ and $V_j$ if $d_{ij}\neq *$.
This fact, together with Lemma \ref{lem:last} suggest the following algorithm
to construct a graph $G\in\langle \mathbb V, d, D^*\rangle$, if one exists:
Let $D_0$ be the matrix we get if we substitute all $*$'s with 0's.
We first run the construction algorithm of this section 
with input $\mathbb V, D_0$.
This will construct a graph $H$ that has $d_{ij}$ edges between
$V_i$ and $V_j$ if $d_{ij}\neq *$ and 0 otherwise, and satisfies the
balanced degree invariant.
Now, what we need to add to get a $G\in\langle \mathbb V, d, D^*\rangle$
is a graph on $V$ with degree sequence defined by $d(V_i)-\deg_H(u)$
for every $u\in V_i$, where $V_i$--$V_j$ edges are forbidden if $d_{ij}\neq *$.
However, this is reduced to finding a matching in a properly defined graph 
$\cal G$. 

To see this, notice that if we want to construct a graph on vertex set
$[n]$ with a given 
degree sequence $d_1 \geq d_2 \geq \ldots \geq d_n$ and a given set $F$ of forbidden edges,
we can define the graph ${\cal G}\! = \! ({\cal V},{\cal E})$
as follows.
The vertex set is ${\cal V} \! = \! \cup_{i=1}^n {\cal V}_i$,
where ${\cal V}_i \! = \! \cup_{j\in [n], j\neq i} \{ v_{ij} \}
                   \cup_{j^{\prime}=1}^{n-d_i-1} \{ u_{ij^{\prime}} \}$,
              $\forall 1 \leq i \leq n$.
The vertices $v_{ij}$ denote a potential
edge between $i$ and $j$ in the graphic realization.
The vertices $u_{ij^{\prime}}$ will enforce the required degrees $d_i$.
Now the edges are
${\cal E} \! = \! \cup_{1 \leq i < j \leq n\:, \:ij\notin F} \{ \{ v_{ij},v_{ji} \} \}
\cup_{i=1}^n \cup_{j=1}^{n-1}
        \cup_{j^{\prime}=1}^{n-d_i-1} \{ \{ v_{ij} , u_{ij^{\prime}} \} \}$.
It is straightforward to verify that
the degree sequence, given $F$, is realizable
if and only if  ${\cal G}$
has a perfect matching.
\end{prf}

%==========================================================
%==========================================================

\section{Joint-Degree Matrix Connected Realization}
\label{sec:algorithm2}

%==========================================================
%==========================================================

We now turn to the question of constructing a connected graphic
realization of an instance $\langle{\mathbb V},d,D\rangle$,
or showing that such a realization does not exist.
It is easy to see that this problem
is different from its counterpart
Erd\H{o}s-Gallai condition (the degrees summing up
to at least $2(n\! - \! 1$)).
In particular, there are graphically realizable instances
of $\langle{\mathbb V},d,D\rangle$ which include many edges,
but have no graphic connected realization
(for example, if all the edges are required to be inside distinct 
classes $V_i$ and $V_j$, $i \! \neq \! j$).
It is also easy to see that arbitrary simple flips
cannot be used to decrease the number of connected components
of a non-connected graphic realization $G \in \langle{\mathbb V},d,D\rangle$.
In particular, let $uv$ and $xy$ be edges in $G$,
let $ux$ and $vy$ be edges not in $G$,
and let $u\in V_i$, $v\in V_j$, $x \in V_{i^{\prime}}$
and $y \in V_{j^{\prime}}$.
Then, the flip of removing $uv$ and $xy$
and adding $ux$ and $vy$ yields
a graph in  $\langle{\mathbb V},d,D\rangle$
if and only if $V_i \! = \! V_{i^{\prime}} $
or/and  $V_j \! = \! V_{j^{\prime}} $.

In what follows we 
give  necessary and sufficient conditions
for $\langle{\mathbb V},d,D\rangle$ to have a connected realization.
The proof provides a polynomial (in $|V|=n$) time algorithm that constructs a
connected realization, if one exists,
or produces a certificate that $\langle{\mathbb V},d,D\rangle$ does not have
a connected realization.

Roughly, the general approach to construct
a connected graph in $\langle{\mathbb V},d,D\rangle$ is to first construct a tree
on $V$
that does not violate the upper bounds specified by $D$.
This will be called a \emph{{valid tree for}} $(\mathbb V, D)$.
If such a tree exists, then Lemma \ref{lem:3-1} shows how to transform it to a tree
that does not violate the upper bounds specified by $D$ and $d$  and also 
satisfies the balanced degree invariant.We call such a tree a \emph{balanced tree for}
$\langle{\mathbb V},d,D\rangle$.
We may then continue with the greedy construction algorithm 
of Section \ref{sec:algorithm1} which,
by Remark 2 at the end of Section \ref{sec:algorithm1}, 
never increases the number of connected components, so that we extend the balanced tree
to a $\langle{\mathbb V},d,D\rangle$ graph.
\begin{lemma}
\label{lem:3-1}
Let $T_{\rm Valid}$ be a valid tree for  $({\mathbb V},D)$.
Then, we can efficiently construct a balanced tree for $\langle{\mathbb V},d,D\rangle$,
 $T_{\rm Balanced}$.
\end{lemma}
\begin{prf}
To construct $T_{\rm Balanced}$ we modify the degrees within each $V_i$. 
We make use of the
fact that, in any tree, if we pick two vertices $u$ and $v$, we can
move any neighbor of $u$ to become a neighbor $v$, with the exception
of the neighbor that lies on the unique $u$--$v$ path of the tree. Let
${\delta}_i$ be the average degree in $T_{\rm Valid}$ of the vertices in
$V_i$. 
Then, as long as
$\exists\:u,v\in V_i$ such that 
${\rm deg}_{T_{\rm Valid}}(u)>\lceil {\delta}_i
\rceil$ and ${\rm deg}_{T_{\rm Valid}}(v) < \lfloor {\delta}_i \rfloor$,
move neighbors of $u$ to $v$, until either ${\rm deg}_{T_{\rm Valid}}(u)=\lceil
{\delta}_i \rceil$, or
${\rm deg}_{T_{\rm Valid}}(v)=\lfloor {\delta}_i \rfloor$. 
This will need at most
$2|V_i|$ iterations. We do this for every $i$ to get a balanced 
tree $T_{\rm Balanced }$.
Notice that, while the degrees are made as equal as possible, the
number of edges between $V_i$ and $V_j$ is not affected, for any $i,j$.
Thus, $T_{\rm Balanced}$ is a balanced tree for $\langle{\mathbb V},d,D\rangle$.
\end{prf}

If there is no connected realization, then we want to produce a
{\it certificate of non existence} of a valid tree.
In general, it is not clear how to construct efficiently a valid tree
or a certificate of non existence of such a tree.
Indeed, the sufficient and necessary conditions for connectivity
listed below appear to require exponential search.
However, our Valid Tree Construction Algorithm
solves both problems in polynomial time.
With this motivation in mind, we proceed to the technical details.

We need some more notation.
Let $G$ be a connected realization in
$\langle{\mathbb V},d,D\rangle$.
Let $\widetilde{\mathbb V} \! =  \! \{ \widetilde{V}_{1},\ldots , \widetilde{V}_{k} \}$,
where $|\widetilde{V}_{i}| \! = \! \max \{ 1, |V_i| \! - \! d_{ii} \}$
is the minimum possible number of connected components of a subgraph of $G$
induced by $V_i$.
Let $\widetilde{D} \! = \! (\widetilde{d}_{ij})$
be derived from $D$ in the natural way:
$\widetilde{d}_{ii} \! = \! 0$ and
$\widetilde{d}_{{ij}} \! = \! \min \{ |\widetilde{V}_{i}|\cdot|\widetilde{V}_{j}| ,d_{ij} \}$.
\begin{lemma}
\label{lem:3-2}
A valid tree $\widetilde{T}$ for
$( \widetilde{\mathbb V}, \widetilde{D} )$
can be transformed efficiently to a valid tree $T$ for $({\mathbb V},D)$.
\end{lemma}
\begin{prf}
To turn $\widetilde{T}$ to a tree $T$ on
$V$, that satisfies the $d_{ij}$'s as upper bounds 
we stick a path of length $|V_i|-|\widetilde{V}_i|$
on an arbitrary vertex of $\widetilde{V}_i$, so that we get $|V_i|$
vertices. We do this for every $i$.
Now notice that for the resulting tree $T$ we have
that the edges of $T$ inside $V_i$ are 
$|V_i|-|\widetilde{V}_i| - 1$. Recall that,
by definition, $|\widetilde{V}_i|=\max\{1, |V_i|-d_{ii}\}$. Therefore,
the edges of $T$ inside $V_i$ are at most $d_{ii}$. 
Moreover,
those paths inside each $V_i$ were the only thing added to $\widetilde{T}$. 
That is, the number of edges between $V_i$ and $V_j$ are
at most
$\widetilde{d}_{ij}\leq d_{ij}$.
Thus we created a tree $T$ on $V$ that does not violate the
$\widetilde{d}_{ij}$'s, and therefore the $d_{ij}$'s.
\end{prf}

Therefore, Lemmata \ref{lem:3-1} and \ref{lem:3-2} reduced the problem 
to finding sufficient and necessary conditions for the existence
of a valid tree $\widetilde{T}$ for $(\widetilde{\mathbb V},\widetilde{D})$,
an efficient construction
for some $\widetilde{T}$, if it exists,
or a certificate that $\widetilde{T}$ does not exist.

We are ready to state the necessary and sufficient conditions. 
Let ${\cal F} \! = \! \{ \widetilde{V}_{{i_1}}, \widetilde{V}_{{i_2}},
\ldots \widetilde{V}_{{i_{\ell}}} \} \subseteq
\{ \widetilde{V}_{1}, \widetilde{V}_{2}, \ldots , \widetilde{V}_{k}  \}$,
and let
${\cal A} \! = \! \{ {\cal A}_1,  {\cal A}_2, \ldots  {\cal A}_{\lambda} \}$
be a partition of ${\cal F}$. The interpretation is that each ${\cal A}_i$ will collapse to a single vertex.
Define the undirected weighted graph
${\cal G} \! = \! ({\cal V},{\cal E},w)$ as follows:\\
$\bullet$ ${\cal V} \! = \! \{ \alpha_1, \alpha_2, \ldots  \alpha_{\lambda} \}
          \cup_{j \neq i_1,i_2, \ldots , i_{\ell}} \{ u_j \}$,
that is, one vertex $\alpha_i$ for each ${\cal A}_i$ and one vertex $u_j$ for each
                   $\widetilde{V}_{j} \not\in {\cal F}$.\\
$\bullet$ If there exists $\widetilde{V}_{x} \in {\cal A}_i$
and $\widetilde{V}_{y} \in {\cal A}_j$ such that $\widetilde{d}_{{xy}} > 0$, then
$\alpha_i\alpha_j \in {\cal E}$ and
$w(\alpha_i , \alpha_j ) \! = \! 1$.  \\
$\bullet$ If there exists $\widetilde{V}_{j} \not\in {\cal F}$,
and, for some $i$, there exists $\widetilde{V}_{x} \in {\cal A}_i$
such that $\widetilde{d}_{{xj}} > 0$, then $\alpha_i u_j \in {\cal E}$
and $w ( \alpha_i , u_j ) \! = \! \min \{ |\widetilde{V}_{j}|,
         \sum_{x : \widetilde{V}_{x} \in {\cal A}_i} \widetilde{d}_{{xj}} \}$.  \\
$\bullet$ If there exist $\widetilde{V}_{i} \not\in {\cal F}$,
$\widetilde{V}_{j} \not\in {\cal F}$ and $\widetilde{d}_{{ij}} > 0$,
then $u_i u_j \in {\cal E}$ and
$w ( u_i , u_j ) \! = \! \widetilde{d}_{{ij}}$.\smallskip

\noindent
{\it Necessary  and Sufficient Conditions:}
Given a connected $G\in \langle \mathbb V, d, D \rangle$ we can easily get a valid
tree  $\widetilde{T}$ for $(\widetilde{\mathbb V},\widetilde{D})$.
To do so, we collapse each connected component of the subgraph $G_i$ of $G$, induced
by $V_i$, in to a single vertex, and if necessary a few of these vertices together, 
so that the cardinality of the vertices of $G_i$ reduces from $|V_i|$ to $|\widetilde{V}_i|$. 
Then, we delete any loops or multiple vertices. The resulting graph is still connected. We do this 
$\forall i$ and then take a spanning tree of the resulting connected graph. This is a valid
tree  $\widetilde{T}$ for $(\widetilde{\mathbb V},\widetilde{D})$.
Then, the existence of
$\widetilde{T}$ implies 
the following necessary condition for a connected realization to exist:
for every ${\cal F}$ and every ${\cal A}$,
the graph ${\cal G} \! = \! ({\cal V},{\cal E},w)$ is connected
and $\sum_{e\in{\cal E}} w(e) \geq |{\cal A}| +
\sum_{i: \widetilde{V}_{i} \not\in {\cal F}} |\widetilde{V}_{i}| -1 $.

On the other hand, we call a pair $({\cal F}, {\cal A})$  such that
the above necessary condition fails,
a {\it certificate that no connected realization of }
$\langle{\mathbb V},d,D\rangle$ {\it exists}.

Next, we will show that the above stated necessary condition
for a connected realization of $\langle{\mathbb V},d,D\rangle$ to exist
is also sufficient.
In particular, we will prove that the  
{\bf Algorithm Valid Tree Construction} below
either produces a valid tree $\widetilde{T}$ for
$(\widetilde{\mathbb V}, \widetilde{D} )$,
or produces a certificate $({\cal F}, {\cal A})$
that no connected realization of $\langle{\mathbb V},d,D\rangle$ exists.

The construction algorithm is as follows.
Let $\widetilde{V} \! = \! \cup_{{i=1}}^k \widetilde{V_i}$.
The algorithm tries to construct a valid tree on $\widetilde{V}$
by maintaining $|\widetilde{V}| \! - \! 1$ edges which
are valid for $(\widetilde{\mathbb V} , \widetilde{D})$
(the number of edges between each $\widetilde{V}_{i}$
and $\widetilde{V}_{j}$ never exceeds $\widetilde{d}_{{ij}}$),
while at the same time decreasing the number of connected components
by adding and removing edges appropriately.
The main idea is that, {\it if two components cannot be connected
in a trivial way} that maintains validity,
{\it then the $\widetilde{V}_{i}$'s that intersect more than one connected
components play a critical role}.
We constantly try to ``free'' an edge incident to such a
$\widetilde{V}_{i}$ while preserving validity and not increasing the
number of connected components.
In the case that such a $\widetilde{V}_{i}$ intersects a cycle,
this is an easy task. Otherwise, we have to remove all the
$\widetilde{V}_{i}$'s that intersect more than one components,
and try to connect two components in the resulting graph.
We recursively repeat this until we connect something,
and then it is easy to find a sequence of adding and removing edges
that connects two components in the original graph and maintains validity.
If the recursion fails, we 
have a certificate
that no connected realization exists.\medskip

\noindent
{\bf Algorithm Valid Tree Construction}
$(\widetilde{\mathbb V}, ~ \widetilde{D})$\\
{\bf begin}\\
$\widetilde{V} \! = \! \cup_{i=1}^k \widetilde{V}_{i}$;$~$
start by a graph $G_0$ consisting of
$|\widetilde{V}| \! - \! 1$ valid edges over $\widetilde{V}$;\\
$j\! = \! 0$; comment: $j$ is the depth of the recursion;\\ 
$G \! = \! G_0$; comment: $G$ is an auxiliary graph;\\
{\bf while} $G_0$ is not connected\\
$~~~~$ {\bf begin}\\
$~~~~$ $A_j \! = \! \{ v~:v~\mbox{lies in some cycle of}~G_j \}$;\\
$~~~~$ $C_j \! = \! \{ \widetilde{V}_{i}~:~A_j\cap\widetilde{V}_{i} \neq \emptyset \}$;
comment: $\widetilde{V}_{i}$'s intersecting some cycle of $G_j$;\\
$~~~~$ $P_j \! = \! \{ \widetilde{V}_{i}~:~\widetilde{V}_{i}~
       \mbox{intersects at least two connected components of}~G_j \}$;\\
$~~~~$ $Z_j \! = \! \{ e\in G_j~:~
     \mbox{at least one endpoint of $e$ is in some $\widetilde{V}_{i}\in P_j$}
               \}$;\\
$~~~~$ {\bf Case 1: if}
       $\exists$ an edge $e\not\in G_0$
                 connecting two connected components in $G$\\
$~~~~~~~~~~~~~~~~$ without violating upper bounds of $\widetilde{D}$ \\
$~~~~~~~~$ add $e$ to $G$; remove any edge from any cycle of $G$;\\
$~~~~~~~~$ $j \! = \! \max \{ j\! - \! 1 , 0 \}$;
           $G_j \! = \! G \cup P_j \cup Z_j$; $G \! = \! G_j$;\\
$~~~~$ {\bf Case 2: elseif} $C_j \cap P_j \! \neq \! \emptyset$\\
$~~~~~~~~$ pick $u,v$ in some $\widetilde{V}_{i} \in C_j \cap P_j$
           that are not connected in $G_0$ and $u\in A_j \cap \widetilde{V}_{i} $;\\
$~~~~~~~~$ find a neighbor $x$ of $u$ that lies on the same cycle in $G$ ;
           find any neighbor $y$ of $x$ in $G$;\\
$~~~~~~~~$ remove $xu$ and $yv$ from $G$ and add $xv$ and $yu$ to $G$;\\
$~~~~~~~~$ $j \! = \! \max \{ j\! - \! 1 , 0 \}$;
           $G_j \! = \! G \cup P_j \cup Z_j$; $G \! = \! G_j$;\\
$~~~~$ {\bf Case 3: elseif} $P_j \! = \! \emptyset$\\
$~~~~~~~~$ let $C_1, C_2, \ldots , C_{\xi}$
           be the connected components of $G_j$;\\
$~~~~~~~~$ let ${\cal A}_i \! = \! \{ \widetilde{V}_{x}~:~
                \widetilde{V}_{x} \subseteq V(C_i) \}$ for $1 \leq i \leq \xi$;
           let ${\cal F} \! = \! \cup_{i=1}^{\xi} {\cal A}_i$;
           let ${\cal A} \! = \! \{ {\cal A}_1, \ldots , {\cal A}_{\xi} \}$; \\
$~~~~~~~~$ output $({\cal F},{\cal A})$ and terminate;
           comment: {\it found a certificate of non existence};\\
$~~~~$ {\bf Case 4: else} $G_{j+1} \! = \! G_j \setminus P_j$;
                   $G \! = \! G_{j+1}$; $j \! = \! j+1$;\\
{\bf end};\\ 
output $G_0$; comment: {\it found a connected realization};\\
{\bf end}.
\begin{thm}
\label{thm:algorithm2}
Algorithm Valid Tree Construction outputs 
a valid tree for 
$( \widetilde{\mathbb V},  \widetilde{D} )$,
if such a tree exists. 
Otherwise, it outputs a certificate $(\mathcal F, \mathcal A)$ showing that 
no such tree exists.
The algorithm runs in time polynomial in $n$. 
\end{thm}
\begin{prf}
First, notice that we start with a graph $G_0$ 
on $|\widetilde{V}|$ vertices and $|\widetilde{V}|-1$ edges, 
that does not violate any upper bounds imposed by $\widetilde{D}$. 
The algorithm also creates at most $k$ graphs
$G_1,G_2,\ldots ,G_k$, 
such that $G_{i+1}$ is an induced subgraph of $G_i$ 
(and thus of $G_0$ as well) on strictly
less vertices. 
To be precise, $V(G_{i})\setminus V(G_{i+1})$ contains one or more of the 
$\widetilde{V}_j$'s. 
To see this, notice that for $G_{i+1}$ to be constructed 
this must happen in Case 4, 
and then $G_{i+1}$ is the subgraph of $G_i$ induced on 
$V(G_i)\setminus\left(\cup_{\widetilde{V}_j\in \mathcal P_i}\widetilde{V}_{j}\right)$. 
But $\mathcal P_i$ in this case should be nonempty,
or the algorithm would have terminated in Case 3.
Now, for each of these graphs, say $G_j$, we associate four sets: \\
$\mathcal P_j$ is the set of the $\widetilde V_i$'s that intersect 
more than one connected component of $G_j$, \\
$\mathcal Z_j$ is the set of edges that have at least one endpoint 
in some $\widetilde V_i\in \mathcal P_j$,\\
$\mathcal A_j$ is the set of all vertices that belong to some cycle in $G_j$,\\
$\mathcal C_j$ is the set of the $\widetilde V_i$'s 
that intersect some cycle in $G_j$.

Notice that if $G_0$ is not connected, then for any $j$ 
such that a $G_j$ is constructed by the algorithm
we have $\mathcal A_j\neq \emptyset$ 
(and thus $\mathcal C_j\neq \emptyset$ 
as well). 
To see this notice that $G_0$ is either a tree, or contains a cycle.
Moreover, whenever a $G_j$ is created (in Case 4),
its vertex set contains $A_{j-1}$ 
(otherwise, the current iteration would not go further than Case 2). 
That is, if $G_{j-1}$
contains a cycle, so does $G_j$ (if created at all). 
The above also implies that whenever $G_j$ is created, 
$V(G_j)\neq \emptyset$.
As discussed above, the algorithm cannot go to Case 4 for $k$ 
consecutive iterations.
Therefore, within $k$ iterations one of Cases 1,2 or 3 happens. 

Suppose that either Case 1 or Case 2 happen,
i.e., two connected components $C,C'$
of $G=G_j$ become connected to each other without any $\widetilde{d}_{ij}$ 
being violated. 
If $j=0$, the number of connected components of $G_0$ is 
decreased. 
Assume not. 
Notice that $C$ and $C'$ in $G_{j-1}$ must be subgraphs of
the same connected component, 
otherwise they would have been connected 
to each other in an earlier iteration,
when $G$ was still $G_{j-1}$. 
Now, $G_{j-1}$ is updated by adding $\mathcal P_{j-1}$ and $\mathcal Z_{j-1}$
back to $G$ and this becomes the current graph $G$. 
Just for notational convenience, we are going to call this graph 
$G_{j-1}'$ as opposed to the old $G_{j-1}$. 
Notice that $\mathcal P_{j-1}$ is not affected, i.e.,
$\mathcal P_{j-1}'=\mathcal P_{j-1}$. 
The key observation now is that, in $G_{j-1}'$ a new cycle is created, 
containing 
one new edge added in the last iteration, 
as well as some $v$ from some $\widetilde{V}_i\in \mathcal P_{j-1}$ 
(that was on a path connecting $C$ and $C'$ in $G_{j-1}$). 
That is, in the next iteration Case 2 will happen
and the number of components of $G_{j-1}'$ will go down by one as well.
So, we have that if Cases 1 or 2 happen, while $G=G_j$, 
then in the next $j$ iterations
Case 2 will happen. 
This results in decreasing the number 
of connected components of $G_{0}$ by one. 

Now suppose that Case 3 happens.
That is, for some $j$, $\mathcal P_j=\emptyset$ and Cases 1 and 2 fail.
Let $C_1,C_2,\ldots,C_{\xi}$ be the connected components of $G_j$ 
and let $\mathcal A$ and 
$\mathcal F$ be defined as in the algorithm. 
The definition of the
$\mathcal A_i$'s makes sense because, in $G_j$, 
if a vertex $v\in\widetilde{V}_x$ is in a component
$C_i$ then all vertices of $\widetilde{V}_x$ are in $C_i$.
Consider the graph $G_j$ together with 
all vertices and edges removed in the previous $j$ iterations, 
i.e., the current $G_0$. Then, there exists some $i$,
such that all cycles in $G_0$
are contained in the subgraph of $G$ induced by the vertices of $C_i$. 
We claim that
in $G_0$, any edge that can be added without violating the upper bounds 
given by $\widetilde{D}$,
must have both its endpoints in some $V(C_i)$.

To prove the latter claim, 
we show that all other possible cases fail to happen. 
Assume that there exists an edge $uv$ that 
can be added to $G_0$ without violating any constraint, 
such that $u\in V(C_i),v\in V(C_{i'})$, where
$i\neq i'$. But then, 
$uv$ would have been found and added in Case 1 of the current iteration. 
In particular, if $\widetilde{V}_x\in \mathcal A_i$ 
and $\widetilde{V}_y\in \mathcal A_{i'}$, 
then we must have $\widetilde{d}_{xy}=0$ 
or else $C_i$ and $C_{i'}$ would become connected
in Case 1 of the current iteration.
Now, assume
that there exists an edge $uv$ 
that can be added to $G_0$ without violating any constraint, 
such that either $u\in V(C_i),v\in \widetilde{V}_y \in\mathcal P_{i'}$ 
with $i'<j$, or 
$u\in \widetilde{V}_x \in\mathcal P_{i},v\in \widetilde{V}_y 
\in\mathcal P_{i'}$ with $i'\leq i<j$.
This means that a few iterations back,
when $G$ was $G_{i'}$, $uv$ was available also. 
But $\widetilde{V}_y$ intersects at least two connected
components of $G_{i'}$ and therefore, 
$\exists\:v'\in \widetilde{V}_x$ such that $u$ and $v'$ lie in 
different connected components in $G_{i'}$. 
Since adding $uv$ is legal, so is adding $uv'$. 
But then,
$uv'$ would have been added in Case 1 of that iteration. This proves the claim.

By the above claim, for any $\widetilde{V}_x, \widetilde{V}_y$ that are not contained 
in $\mathcal F$,
we must have as many edges as possible between $\widetilde{V}_x$ and
$\widetilde{V}_y$, that is, if $G[ X,Y ]$ 
denotes the bipartite graph induced by $X$ and $Y$,
$$|E(G_0[\widetilde{V}_x,\widetilde{V}_y])|=\widetilde{d}_{xy}.$$ 
Also, for any $\widetilde{V}_x$ and 
$\mathcal A_i$ we must already have 
as many edges as possible between $\widetilde{V}_x$ and
all the $\widetilde{V}_y$'s in $\mathcal A_i$ in $G_0$. That is,
$$|E(G_0[\widetilde{V}_x,\cup_{y:\widetilde{V}_y\in \mathcal A_i}\widetilde{V}_y])|=
\max\{|\widetilde{V}_x|,\sum_{y:\widetilde{V}_y\in \mathcal A_i}\widetilde{d}_{xy}\}.$$ 

Now notice that if we identify all $\mathcal A_i$ 
with one single vertex to get $H$ from
$G_0$, $H$ remains disconnected, 
but contains no cycles. That is, $|E(H)|< |V(H)|-1$. 
But,
$$|V(H)|=|\mathcal A| + \sum_{i: \widetilde{V}_i\notin \mathcal F}|\widetilde{V}_i|, \text{ and}$$
$$|E(H)|=\frac{1}{2}\sum_{x: \widetilde{V}_x\notin \mathcal F} 
\sum_{y: \widetilde{V}_y\notin \mathcal F} 
\widetilde{d}_{xy} + \sum_{x: \widetilde{V}_x\notin \mathcal F}
\sum_{i=1}^{\xi}\max\{|\widetilde{V}_x|, \sum_{y:\widetilde{V}_y\in 
\mathcal A_i}\widetilde{d}_{xy}\}$$
That is, if we use $\mathcal F$ and $\mathcal A$ 
to construct the weighted graph 
$\mathcal G= (\mathcal V, \mathcal E, w)$ as in the necessary and sufficient condition, then 
$$\sum_{e\in\mathcal E} w(e) <
|\mathcal A | + 
\sum_{i: \widetilde{V}_i\notin \mathcal F} |\widetilde{V}_i| - 1,$$
i.e., the condition fails to hold and $(\mathcal F, \mathcal A)$ is indeed a 
certificate showing that no connected graph in 
$\langle{\mathbb V},d,D\rangle$ exists.

From the above, within no more than $2k$ iterations, 
Algorithm Valid Tree Construction either reduces the 
number of components of $G_0$, or terminates, 
giving a certificate $(\mathcal F, \mathcal A)$. 
Therefore, in at most
$2k|\widetilde{V}|$ iterations the algorithm terminates, and if
a valid tree $\widetilde{T}$ of $(\mathbb{\widetilde{V}}, \widetilde D )$  exists, 
$G_0$ is going to be such a tree.
\end{prf}

%==========================================================
%==========================================================

\section{A step towards sampling}
\label{sec:Sampling}

%==========================================================
%==========================================================

\begin{figure}[h]
\hspace{-0.5cm}
\begin{minipage}[b]{0.65\linewidth}
\centering
\includegraphics[scale=0.6]{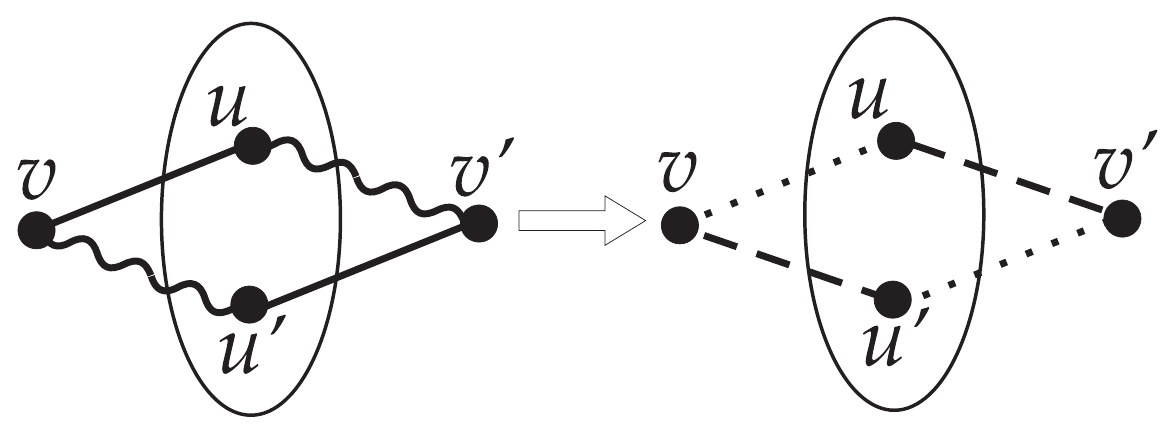}
\caption{A legal switch.}
\label{fig1}
\end{minipage}
\hspace{-3cm}
\begin{minipage}[b]{0.65\linewidth}
\centering
\includegraphics[scale=0.6]{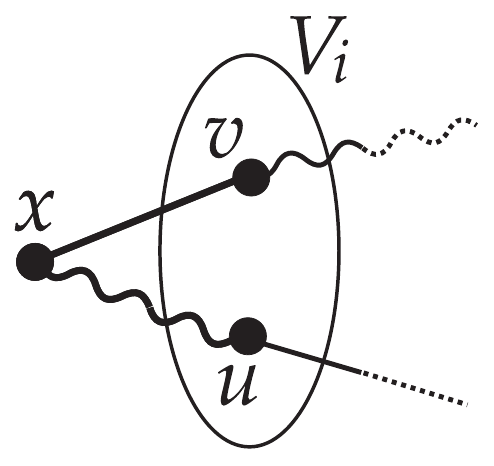}
\caption{A pairing node.}
\label{fig2}
\end{minipage}
\end{figure}

Towards uniform sampling from $\Omega=\langle{\mathbb V},d,D\rangle$,we can define 
a natural Markov chain on $\Omega$.
Let $G\in \Omega$ and $u,u'$ be two vertices that belong to $V_i$ for some $i$.
Also let $v,v'$ be vertices such that $uv,u'v' \in E(G)$, but $uv', u'v \notin E(G)$.
Then, the graph $G'$ with $E(G')=E(G)\cup \{uv',u'v\} \setminus \{uv,u'v'\}$ is 
still in $\Omega$. We call such an operation a \emph{legal switch} and denote it
as $[uv,u'v'|uv',u'v]$ (Figure \ref{fig1}). Two legal switches are distinct, if they produce different
graphs.\medskip

We define the Markov chain $\mathcal M$ as follows.
Given a state/graph $X_t=G\in\Omega$, first calculate the number of distinct legal 
switches, $\ell(G)$. Notice that this can be done in $O(n^4)$ time.
\begin{itemize}
	\item W.p. $1/2$, let $X_{t+1}=G$.
	\item W.p. $1/2$, choose one of the $\ell(G)$ distinct legal switches. Perform
	the switch and let $G'$ be the resulting graph. With probability
	$\frac{\ell(G)}{\ell(G)+\ell(G')}$ let $X_{t+1}=G'$, otherwise let $X_{t+1}=G$.
\end{itemize}

Clearly, $\mathcal M$ is aperiodic. Notice that if $\pi$ is the uniform distribution on $\Omega$,
for all $x, y \in \Omega, \pi(x)P(x, y) = \pi(y)P(y, x)$. That is, if $\mathcal M$ is
irreducible, then its unique stationary distribution is the uniform distribution $\pi$.

Below we show that $\mathcal M$  is irreducible. Usually this is a trivial step,
but here it is a bit more involved. We should note here that 
\cite{SP12} independently proposed an alternative proof for the irreducibility of $\mathcal M$. This proof however was flawed, as noted in \cite{CDEM15}, where an alternative proof is given. Our approach below is simpler, although not completely straightforward.  Let $G_0$ and
$G_1$ be two arbitrary instances in $\Omega$. We want to show that there exists
a sequence of legal switches that applied to $G_0$ gives $G_1$. First, we need 
to introduce some notation and terminology. In what follows $X=G\oplus G_1$ is the symmetric
difference of the current graph $G$ and $G_1$; initially, $G=G_0$. We will refer 
to the edges of $E(G)\setminus E(G_1)$ as \emph{straight}, and the edges of 
$E(G_1)\setminus E(G)$ as \emph{squiggly}. The graph $X$ is the union of straight 
and squiggly edges. Also, we will refer to the edges of $E(G)\cap E(G_1)$ as
\emph{dashed}, and to the rest of the edges (edges in neither $G$ nor $G_1$) 
as \emph{dotted}. Obviously, the depiction of the edges will reflect their names.

If there exist vertices $x,u,v$ of $X$ such that $xu$ is straight, $xv$ is squiggly
and $u,v$ belong to the same $V_i$, then we call $x$ a \emph{pairing node} (Figure \ref{fig2}).

Finally, notice that $\forall v\in X$ the number of straight and the number of
squiggly edges adjacent to $v$ are equal, i.e., they are ${deg_X(v)}/{2}$.

\begin{lemma}
The Markov chain $\mathcal M$ defined above is irreducible.
\end{lemma}

\begin{prf}
By induction on $|E(X)|$. If $|E(X)|=4$, then $X$ has to contain an alternating 
cycle of length 4, with straight and squiggly edges, with two non adjacent vertices 
in the same $V_i$. Thus we can go from $G$ to $G_1$ in one legal switch.

Now assume that $|E(X)|= 2k>4$. 

\noindent\textbf{Case 1:} $X$ contains a pairing node $x$.

\noindent\emph{Subcase 1a:} There exists a straight neighbor of $u$, say $w$, that 
is also a squiggly neighbor of $v$ (Figure \ref{fig3}a). Then, by switching $[vx,uw|vw,ux]$ in $G$, 
we reduce $|E(X)|$ to $2k-4$. Thus, by induction, we can go from $G$ to $G_1$ by 
performing a sequence of legal switches.

\noindent\emph{Subcase 1b:} There exists a straight neighbor of $u$, say $w$, that 
is also a dotted neighbor of $v$, or there exists a dashed neighbor of $u$, $w$, that 
is also a squiggly neighbor of $v$ (Figure \ref{fig3}b). Then, by switching $[vx,uw|vw,ux]$ in $G$, 
we reduce $|E(X)|$ to $2k-2$. Thus, by induction, we can go from $G$ to $G_1$ by 
performing a sequence of legal switches.

\noindent\emph{Subcase 1c:} There exists a straight neighbor of $u$, say $w$, that 
is also a dashed neighbor of $v$, or there exists a dotted neighbor of $u$, $w$, that 
is also a squiggly neighbor of $v$ (Figure \ref{fig3}c). Notice that by switching $[ux,vw|uw,vx]$ 
in $G_1$, we get a graph $G_1'$, such that $|E(G\oplus G_1')|=2k-2$. Thus, by induction, 
we can go from $G$ to $G_1'$ by performing a sequence of legal switches. Then, by 
switching $[ux,vw|uw,vx]$ in $G_1'$, we get $G_1$.

\begin{figure}[h]
\centering
\includegraphics[scale=0.55]{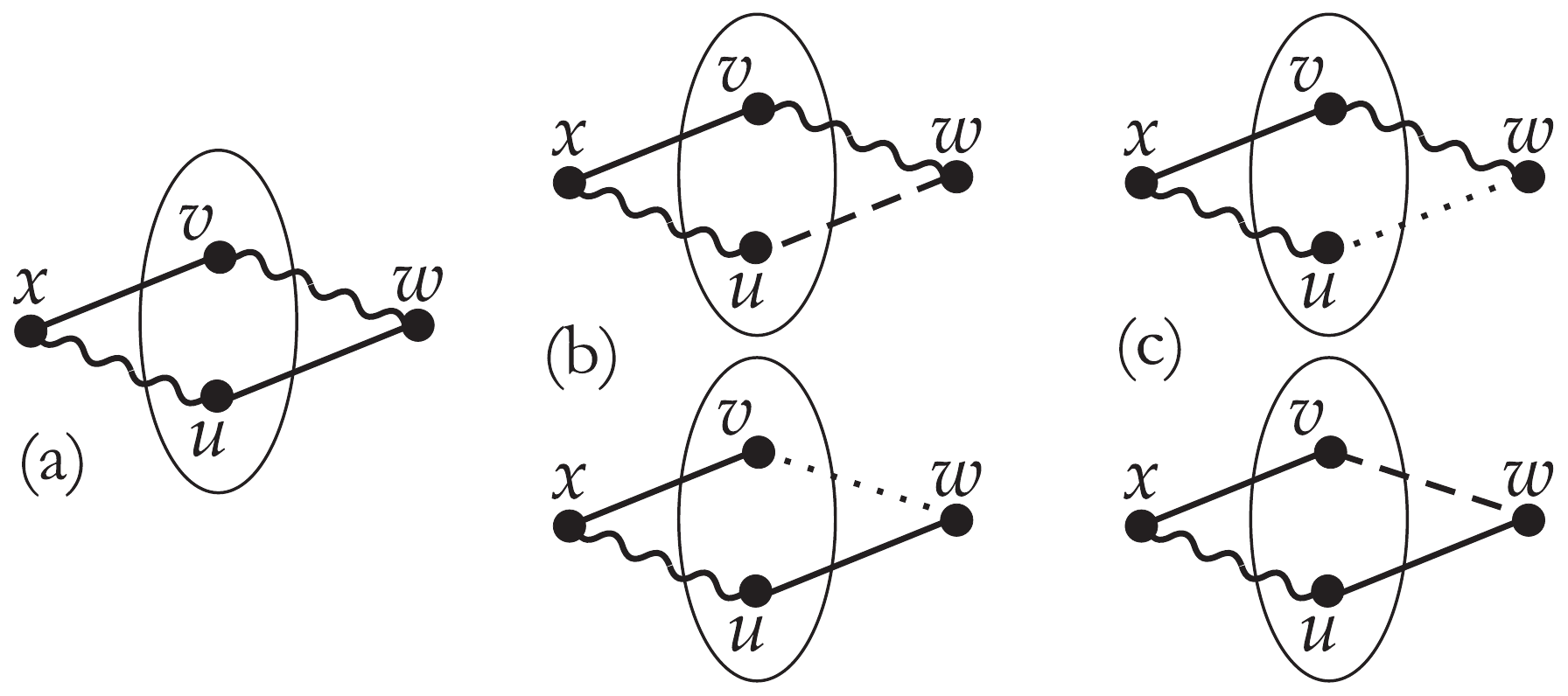}
\caption{Case 1.}
\label{fig3}
\end{figure}

Now, we claim that always one of the above subcases holds. Assume not. That is, \emph{(i)} 
all straight neighbors of $u$ ($v$ excluded) are also straight neighbors of $v$ and 
\emph{(ii)} all squiggly neighbors of $v$ ($u$ excluded) are also squiggly neighbors 
of $u$. Since, $x$ is a squiggly neighbor of $u$ and a straight neighbor of $v$,
\emph{(i)} implies that $deg_X(u)/2<deg_X(v)/2$ (Figure \ref{fig4}a) and \emph{(ii)} implies 
that $deg_X(u)/2>deg_X(v)/2$ (Figure \ref{fig4}b), producing a contradiction.\medskip

\begin{figure}[h]
\centering
\includegraphics[scale=0.65]{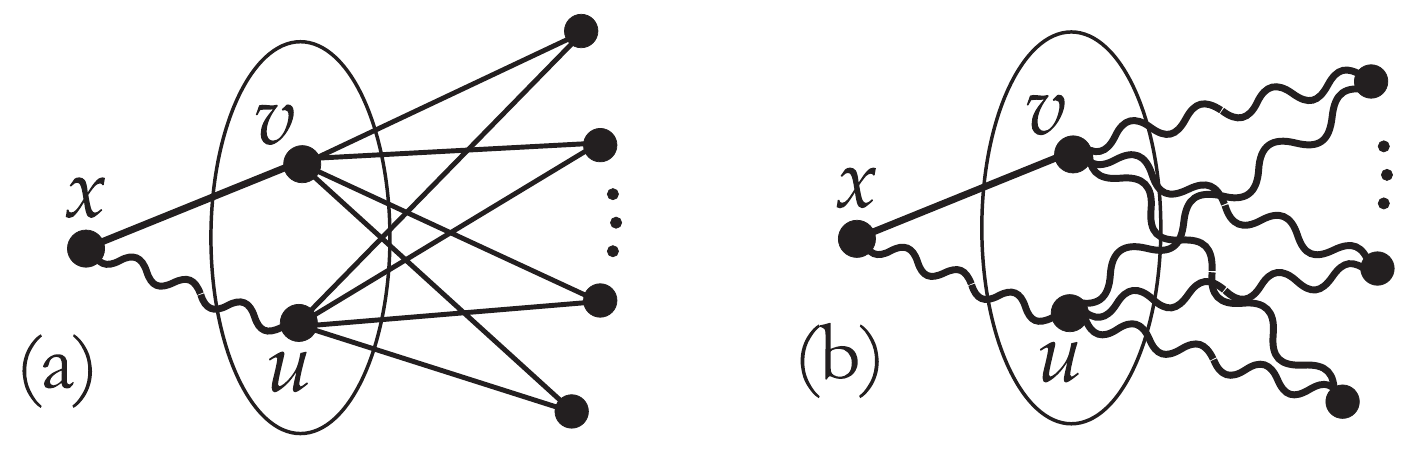}
\caption{Subcases 1a, 1b, and 1c exhaust all possibilities for Case 1.}
\label{fig4}
\end{figure}

\noindent\textbf{Case 2:} $X$ contains no pairing node. 

\noindent Then, there exist two edges, one squiggly $xu$ and one straight $yv$ such 
that $u,v,x,y$ are all distinct and $x,y\in V_i$, $u,v\in V_j$ for some $i,j$ (Figure \ref{fig5}).
\begin{figure}[h]
\centering
\includegraphics[scale=0.65]{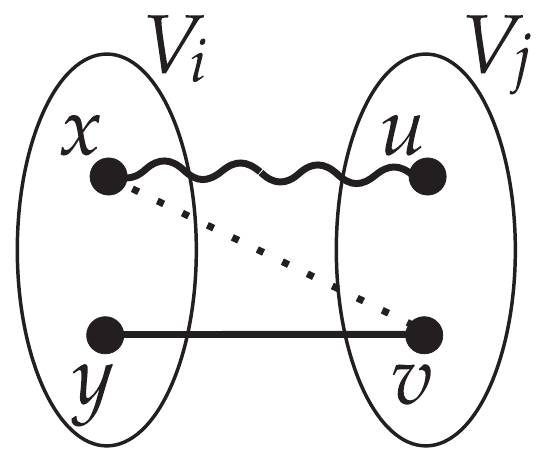}
\caption{If no pairing node exists.}
\label{fig5}
\end{figure}
Notice that $xv$ can only be either dashed or dotted. Assume it is dotted. (\emph{The case 
where $xv$ is dashed is symmetric}.) Then, $v$ is a neighbor of $y$ in $G$, but not 
of $x$. Since $x$ and $y$ have the same degree in $G$, there exists some $w$ such that 
$xw$ is in $E(G)$, while $yw$ is not.

\noindent\emph{Subcase 2a:} The edge $xw$ is straight and $yw$ is dotted. Then, by 
switching $[xw,yv|xv,yw]$ in $G$, we keep $|E(X)|= 2k$ and we create the pairing node 
$x$ (Figure \ref{fig6}a). By Case 1 above and induction, we can go from $G$ to $G_1$ by 
performing a sequence of legal switches.

\noindent\emph{Subcase 2b:} The edge $xw$ is dashed and $yw$ is squiggly. Like above, 
by switching $[xw,yv|xv,yw]$ in $G$, we keep $|E(X)|= 2k$ and we create the pairing 
node $x$ (Figure \ref{fig6}b). By Case 1 above and induction, we can go from $G$ to $G_1$ by 
performing a sequence of legal switches.

\noindent\emph{Subcase 2c:} The edge $xw$ is dashed and $yw$ is dotted. Then, by 
switching $[xw,yv|xv,yw]$ in $G$, we have $|E(X)|= 2k+2$ and we create two pairing nodes 
$x,w$ (Figure \ref{fig6}c). By examining the pairing node $x$ with $u,v$ like we did in subcases 1a, 
1b, 1c above, we either reduce $|E(X)|$ to $2k-2$ and proceed with induction, or we
reduce $|E(X)|$ to $2k$, while $X$ has the pairing node $w$, in which case, by Case 1 
above and induction, we can go from $G$ to $G_1$ by performing a sequence of legal switches.

\begin{figure}[h]
\centering
\includegraphics[scale=0.55]{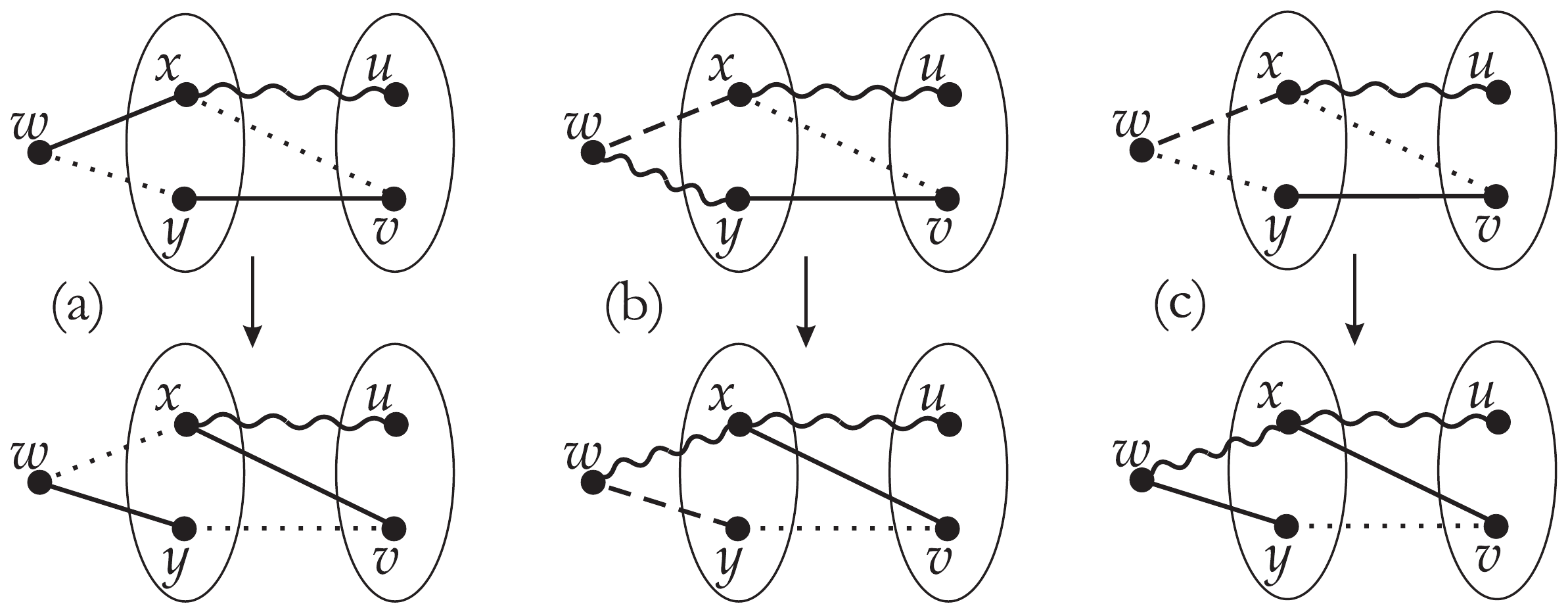}
\caption{Subcases 2a, 2b, and 2c}
\label{fig6}
\end{figure}

Notice that in Case 2, it is not possible to have $xw$ to be straight and $yw$ to be
squiggly.\\

\end{prf}

%\begin{figure}[p]
%\centering
%\begin{minipage}[b]{0.5\linewidth}
%\centering
%\includegraphics[scale=0.6]{fig1}
%\caption{A legal switch.}
%\label{fig1}
%\end{minipage}
%\hspace{0.5cm}
%\begin{minipage}[b]{0.5\linewidth}
%\centering
%\includegraphics[scale=0.6]{fig2}
%\caption{A pairing node.}
%\label{fig2}
%\end{minipage}
%\includegraphics[scale=0.5]{fig3}
%\caption{Case 1.}
%\label{fig3}
%\includegraphics[scale=0.6]{fig4}
%\caption{Subcases 1a, 1b, and 1c exhaust all possibilities for Case 1.}
%\label{fig4}
%\includegraphics[scale=0.6]{fig5}
%\caption{If no pairing node exists.}
%\label{fig5}
%\includegraphics[scale=0.5]{fig6}
%\caption{Subcases 2a, 2b, and 2c}
%\label{fig6}
%\end{figure}

%==========================================================
%==========================================================

\vspace{-3 ex}\section{Further Directions}
\label{sec:discussion}

%==========================================================
%==========================================================

For the original 
degree sequence graphic realization problem,
when a graphic realization of a degree sequence 
$d_1 \geq d_2 \geq \ldots \geq d_n$ exists,
several interesting generalizations
can be solved efficiently.
For example, 
if there are costs on edges, then we
can find a minimum cost realization and 
we can generate a realization, or a connected realization 
uniformly at random 
(under mild restrictions on the degree sequence \cite{JS2,ALENEX03,SODAMCMC,FOCS06}).

The above problems have efficient algorithms
because the original problem has a reduction
to perfect matchings \cite{Val79,JS2,Sinclair}.
In particular, for
$d_1 \geq d_2 \geq \ldots \geq d_n$ as above,
define the graph ${\cal G}\! = \! ({\cal V},{\cal E})$
as follows.
The vertex set is ${\cal V} \! = \! \cup_{i=1}^n V_i$,
where $V_i \! = \! \cup_{j\in [n], j\neq i} \{ v_{ij} \}
                   \cup_{j^{\prime}=1}^{n-d_i-1} \{ u_{ij^{\prime}} \}$,
              $\forall 1 \leq i \leq n$.
The vertices $v_{ij}$ denote a potential
edge between $i$ and $j$ in the graphic realization.
The vertices $u_{ij^{\prime}}$ will enforce the required degrees $d_i$.
Now the edges are
${\cal E} \! = \! \cup_{1 \leq i < j \leq n} \{ \{ v_{ij},v_{ji} \} \}
\cup_{i=1}^n \cup_{j=1}^{n-1}
        \cup_{j^{\prime}=1}^{n-d_i-1} \{ \{ v_{ij} , u_{ij^{\prime}} \} \}$.
It is straightforward to verify that
the degree sequence is realizable
if and only if  ${\cal G}$
has a perfect matching.

Is there a reduction from the
joint-degree matrix realization problem to some
version of matching, flow, or a similar
better understood combinatorial problem?
The relatively smooth decision and construction
algorithm outlined in Section \ref{sec:algorithm1} for realizability
suggests that such a reduction might exist.
One should probably try to reverse engineer the construction algorithm;
however, the main difficulty is outlined in Remark 1 at the end of 
Section \ref{sec:algorithm1},
namely, the alternating sequences of edges involved in the
algorithm are not pure augmenting paths. The existence of a
such a reduction may help solve the problems listed in the next paragraph.

The most interesting open question is undoubtedly whether one can efficiently sample from $\langle{\mathbb V},d,D\rangle$. It would be interesting to 
prove that the Markov chain we suggest in Section \ref{sec:Sampling}, or any other
Markov chain for that matter, is rapid mixing. Although some progress has been made in this direction in \cite{EMT15}, the problem remains largely open. Similar questions can be asked for a weighted version of the problem.
Let $\langle{\mathbb V},d,D\rangle$ be an instance of the joint-degree matrix
realization problem.
If there is a cost associated with every potential edge,
can we construct a realization of minimum cost?
More importantly, can we generate uniformly at random such a realization?

It is also natural to define the following generalization of the graphic realization problem, where conditions on the number of edges
involve arbitrary subsets of vertices.
In particular, for $V= [n]$
and positive integers $d_1 \geq d_2 \geq \ldots \geq d_n$,
let ${\mathbb S}= \{ S_1, S_2 , \ldots , S_k \}$
be an arbitrary partition of $V$,
and let $D\! = \! (d_{ij})$, $1\leq i \leq j \leq k $,
be a $k\times k$ matrix, where $d_{ij}$ is the number of edges between $S_i$ and $S_j$. As is, this is the same as the \emph{partition adjacency matrix problem} introduced in \cite{Ch14}, and the \emph{skeleton graph problem} introduced in \cite{EHIM15}, where some special cases are studied.
Is there a polynomial decision/construction algorithm for the above problem in general?
Notice that this is not a direct generalization of the joint-degree matrix problem, as was noted in \cite{EHIM15}, unless we modify the above definition to specify the degree subsequences in each $S_i$.

%\newpage

%%%%%%%%%%%%%%%BIBLIOGRAPHY BEGINS%%%%%%%%%%%%%%%%%%%%%%%%%%%%%%%%%

%%{\small
%\bibliography{my}
\bibliographystyle{plain}
%%}

%%%%%%%%%%%%%%%BIBLIOGRAPHY ENDS%%%%%%%%%%%%%%%%%%%%%%%%%%%%%%%%%%%

\end{document}